\documentclass[a4paper,12pt]{article}
\usepackage{sectsty}
\usepackage{setspace}
\usepackage{graphicx}
\usepackage{amsmath, amsthm, amssymb}
\usepackage{enumitem}
\usepackage[margin=20mm]{geometry}
\usepackage{natbib}
\usepackage{hyperref}
\usepackage{authblk}
\usepackage[version=4]{mhchem}
\usepackage{lmodern}
\usepackage{txfonts}
\usepackage{cleveref}
\usepackage[table,usenames,dvipsnames]{xcolor}
\usepackage{comment}
\definecolor{darkblue}{rgb}{0.0,0.0,0.3}
\hypersetup{colorlinks,breaklinks,linkcolor=darkblue,urlcolor=darkblue,anchorcolor=darkblue,citecolor=darkblue}

\usepackage[justification=centering]{caption}
\usepackage[labelformat=simple]{subcaption}

\usepackage{ifthen}
\def\eprinttmp@#1arXiv:#2 [#3]#4@{\ifthenelse{\equal{#3}{}}{\href{http://arxiv.org/abs/#1}{arXiv:#1}}{\href{http://arxiv.org/abs/#2}{arXiv:#2 [#3]}}}
\newcommand{\eprint}[1]{\eprinttmp@#1arXiv: []@}
\newcommand{\doi}[1]{\href{http://dx.doi.org/#1}{doi:#1}}

\sectionfont{\normalfont\large\bfseries}
\subsectionfont{\normalfont\normalsize\bfseries}
\subsubsectionfont{\normalfont\normalsize\scshape}

\makeatletter
\g@addto@macro\quote{\small\singlespacing\upshape\sffamily\vspace{-4mm}}
\makeatother

\usepackage{booktabs}
\usepackage{tikz}
\usetikzlibrary{calc,arrows,decorations.markings,shapes,positioning}

\usepackage{multirow}

\newcommand{\quics}{\emph{Joint Center for Quantum Information and Computer Science (QuICS)}, University of Maryland and NIST}

\usepackage[version=4]{mhchem}
\setlist[description]{font={\itshape}}

\begin{document}

\sloppy

\title{How to engineer a quantum wavefunction}

\author[1]{\bf Peter W. Evans\thanks{email: \href{mailto:p.evans@uq.edu.au}{p.evans@uq.edu.au}}}
\author[2]{\bf Dominik Hangleiter\thanks{email: \href{mailto:dhang@umd.edu}{dhang@umd.edu}}}
\author[3]{\bf Karim P. Y. Th\'ebault\thanks{email: \href{mailto:karim.thebault@bristol.ac.uk}{karim.thebault@bristol.ac.uk}}}

\affil[1]{\small{{\it School of Historical and Philosophical Inquiry}, University of Queensland}}
\affil[2]{\small{\quics}}
\affil[3]{\small{{\it Department of Philosophy}, University of Bristol}}

\date{}

\maketitle

\vspace{-10mm}

\begin{abstract}
  In a conventional experiment, scientists typically aim to learn about target systems by manipulating source systems of the same \emph{material type}. In an analogue quantum simulation, by contrast, scientists typically aim to learn about target quantum systems of one material type via an experiment on a source quantum system of a different material type. In this paper, we argue that such inferences can be justified by reference to source and target quantum systems being of the same \emph{empirical type}. We illustrate this novel experimental practice of \emph{wavefunction engineering} with reference to the example of Bose-Hubbard systems.
\end{abstract}

 \tableofcontents

\section{Introduction}

Modern experimental practice allows for a staggering degree of control over lab-based quantum systems. This high level of control operates in terms of both the precision with which quantum systems can be probed and the range of scales of components that can be manipulated: from thousands of ultracold atoms controlled using arrays of laser beams to individual ions that can be electronically trapped. The potential implications of such quantum technology are powerful, wide-ranging, and radical. In this paper our focus is on the particular context of \emph{analogue quantum simulation} in which a well-controlled quantum system in the lab is specifically deployed by scientists to \emph{learn} about features of another quantum system to which they do not have direct access \citep{Dardashti:2015,Dardashti:2019,Thebault:2016,crowther:2019,EvansThebault2020,HangleiterCarolanThebault2017,Field:2022,bartha_2022}.

On what basis should we categorise different physical systems as tokens of the same type? One option is to distinguish types of physical systems by their material constitution, focusing on properties such as masses, atomic constitution, geometry, charges, interactions and the like that designate the detailed physical properties of the system. Call this the \emph{material type} view. The second option is to distinguish types of physical systems by structural similarity in empirical behaviour. In particular, we could take any two physical systems to be of the same type when in some specified parameter regime a set of experimental prescriptions result in appropriately similar measurement outcomes. Call this the \emph{empirical type} view.

The relevance of the distinction between material-type and empirical-type views arises in the context of analogue experiments wherein a \emph{source} system is manipulated in the lab with the aim of gaining understanding of a \emph{target} system which is not directly manipulated. Significantly, the form of justification for the source--target inferences involved in analogue simulation is sensitive to how widely we draw the category of types of physical systems. When the material-type view is assumed, we find that analogue simulations by definition involve a novel form of inter-type uniformity reasoning requiring justification by way of `universality' arguments.\footnote{The sense of universality here is a broad one: two systems may be of the same `universality type' in this sense without being in the same `universality class' in the Wilsonian sense, c.f. \citep{batterman:2019,Gryb:2021}.} However, by contrast, when the empirical-type view is assumed, a more conventional form of intra-type uniformity reasoning is applied, albeit with an atypical notion of type.

The key distinction between these two ways of reconstructing the inferential practices underpinning analogue simulation is that they lead to differences in the conditions that limit the \emph{strength of inductive support} for conclusions about the target system based upon experimental manipulation of the source system. In particular, in the context of the material-type view, and any associated inter-type uniformity reasoning requiring justification by way of universality arguments, \citet{Field:2022} has convincingly argued that inferentially strong conclusions require either (1) detailed knowledge of the microstructure of the source and target, or (2) empirical evidence for the applicability of relevant universality arguments via empirical access to the macro-behaviour of the source and target. Correspondingly, in the context of the empirical-type view, on our account, one is licensed to draw inferentially strong conclusions regarding the target system in a context where we have (i) empirical evidence of the validity of the respective models via detailed knowledge of the microstructure of the source and target, and (ii) empirical evidence for membership of the same empirical type via empirical access to the macro-behaviour of the source and target.

In each case a contrast in terms of strength of inference can be made. On the one hand, in exotic examples of analogue simulation, such as analogue Hawking radiation, conditions such as (1) and (2) can be expected to fail, since for the target system we have the combination of inaccessibility and lack of reliable theories of microstructure. On the other hand, even in the case where the conditions do obtain, we do not generically expect our inferences to meet the gold evidential standards found in systematic, direct experimentation. The empirical-type view, therefore, allows us to understand how a moderate level of evidential support can accrue for hypotheses regarding a target system in an analogue simulation in contexts where the target system is experimentally accessible in some regime, and thus where the reliability of models of the target system, in some parameter regime, can be established. The aim of the analogue simulation is thus to probe the behaviour of an accessible system in an inaccessible regime based upon manipulation of a further system of the same empirical type. Whilst such a pattern of inference is implicit in a wide range of scientific discussions, it has as yet not been subject to explicit philosophical analysis.

In this paper, we argue that the scientific practice of analogue quantum simulation provides a compelling example in which the empirical-type view allows for inductive arguments towards inferences about the behaviour of an accessible target system in an inaccessible regime. 
The fact that both systems are adequately modelled within the framework of quantum theory allows us to run a `bootstrapping' inference wherein the 
general empirical support for the `quantumness' of source and target is combined with direct  empirical evidence of the applicability of an idealised quantum model to the target in an accessible regime, towards the inference of applicability of the target model to a broader inaccessible regime. The two ingredients in the justification of this bootstrapping thus directly correspond to specific realisations of (i) and (ii).
First, we have a specific, empirical premise based upon the experimental manipulation of the target system in the accessible regime, labelled (H) below. Second, we have a broad, theoretical-empirical premise based upon the assumed applicability of quantum theory to the target system, labelled (Q) below.

Through (Q) our argument pattern makes crucial use of quantum theory as a generalised framework that underlies the modelling of quantum systems. 
It is for this reason that we characterise the relevant experimental practice as \emph{wavefunction engineering}.  
Furthermore, our argument employs a quantum uniformity principle, which can be understood as a meta-principle for this kind of modelling practice. At a high level, our argument makes use of the de-idealisation of a single idealised quantum model to both source and target system models and in so doing provides justification for reasoning based upon regularity within empirical types. 
While we do not claim that such empirical-type-regularity based reasoning renders inferences about the target system in an analogue quantum simulation on a par with inferences in the context of conventional experiments, we do argue that the inferences in analogue quantum simulations command stronger inductive support than those in which the target system is inaccessible and the relevant target system model subject to entirely theoretical support.

\section{A Case Study of Analogue Quantum Simulation: Bose-Hubbard Physics}
\label{sec:case study}

Successful analogue quantum simulation requires that the source and target system models are de-idealisations of a single theoretical model in some appropriate regime of idealisation \citep{HangleiterCarolanThebault2017}. The key to the simulation is that the source system can be controlled more easily than the target system, and so an experiment on the source system can probe elements of the target system that are experimentally inaccessible, given that the idealised model is appropriately verified.

One class of dynamical systems that are particularly ripe for modelling in analogue quantum simulation experiments are those that conform to the Bose-Hubbard model, which describes the dynamics of a lattice of interacting bosons. The Bose-Hubbard model was first derived by \citet{Gersch1963} in the context of granular superconductors---a special case of so-called `type-II' superconductors. However, it was the discovery of \emph{quantum phase transitions} at zero temperature between a superconducting and an insulating phase in granular superconductors that sparked theoretical and experimental interest in the model \citep[566]{Bruder2005}. This led to the experimental investigation of other systems described by the Bose-Hubbard Hamiltonian, including thin Helium-films and arrays of superconductors connected by Josephson junction.

Such implementations of the Bose-Hubbard model are engineered systems with extraordinary phase behaviour that display close similarity to the behaviour of `natural' type-II superconductors. Type-II superconductors are characterised by an atypical intermediate phase between their insulating and superconducting phases, an analogue of which is observed in the behaviour of thin Helium-films, and by the formation of magnetic vortices when an external magnetic field is applied, which is observed in Josephson junction arrays. A precise understanding of superconductors is crucial for a broad range of technological applications.

Remarkably, it was found that bosonic atoms loaded into an optical lattice potential created using laser light are also described by the Bose-Hubbard model \citep{Jaksch1998}. The experimental accessibility of this system allows a great range of experimental investigations of Bose-Hubbard dynamics that is inaccessible by other means. 
The potential of cold atoms as an analogue simulation platform was experimentally actualised with the observation that they undergo the same phase transition at zero temperature between a superfluid and an insulator phase \citep{greiner_quantum_2002}. 
The phase transitions in these very different systems are underpinned by the same physical principles: 
that is, the ``competition between the trend to global coherence, due to the hopping of bosonic particles, and the tendency towards localization induced by the strong interactions'' \citep{Bruder2005}.

More specifically, the Bose-Hubbard model is characterised by the Hamiltonian
\begin{equation}
    \label{eq:bose hubbard}
    H_{\text{BH}} = -J \sum_{\langle j,k \rangle} \left(b^{\dagger}_{j}b_{k} + b^{\dagger}_{k}b_{j}\right) + U \sum_{j}b^{\dagger}_{j}b^{\dagger}_{j}b_{j}b_{j} + \sum_{j}\mu_{j}b^{\dagger}_{j}b_{j}~,
\end{equation}
where the bosonic creation and annihilation operators $b^{\dagger}_{j}$ and $b_{j}$ represent atoms at lattice site $j$, and the different terms represent the energy gain $J$ when atoms hop between neighbouring sites, the energy cost $U$ of two atoms at the same site, and the energy offset $\mu_{j}$ of each lattice site. 
Zero-temperature or \emph{quantum} phase transitions can be understood as the transition between regimes in which one of $J$ or $U$ dominates the ground state of the model. 
When $J$ dominates, hopping behaviour is much more likely to occur, and so the ground state consists of delocalised bosons across the lattice. This is the superfluid phase. When $U$ dominates, there is a strong local repulsion between atoms occupying the same lattice site that prevents global coherence. This is the Mott insulator phase \citep[567]{Bruder2005}.

Cold-atom bosonic systems in an optical lattice are accessible to experimental manipulation and probing of a sort not possible for its potential target systems. A cold-atom system is typically constructed by employing counter-propagating lasers combined with a magneto-optical trap to form a space-dependent lattice potential, which is used as a location grid in which ultracold atoms, such as \ce{^{87}Rb}, can be positioned.  This system is accurately described by the Bose-Hubbard Hamiltonian \citep{Jaksch1998}
in a parameter regime where: (i) next-nearest neighbour hopping and nearest-neighbour repulsion are negligible; (ii) the spatial extent of the wavefunction of each oscillator ground state matches the dimensions of the lattice wells; and (iii) the onsite interaction strength is sufficiently small for the number of particles per site. Importantly, all the model parameters can be manipulated by varying an external magnetic field and the amplitude and phase of the lasers generating the lattice potential \citep[33]{HangleiterCarolanThebault2017}. Thus, the zero-temperature phase transition of the system can be controlled. 
What is more, location and momentum information of the atoms in the lattice can be measured with remarkable precision \citep{Bruder2005}.

\subsection{The Target Systems}
\label{sec:analogue bh systems}

We present here three potential target systems for analogue quantum simulation, the set of which demonstrates the variety of material physical systems that can be targeted by the cold-atom source system.

\subsubsection{Superfluid \texorpdfstring{\ce{^{4}He}}{4He} in Vycor}

Vycor is a specially manufactured high-silica glass. When manufactured as a porous structure, it is an ideal substrate for the study of confined liquids in condensed matter physics. Helium-4 adsorbed in Vycor is observed to form a superfluid: it behaves as an interacting ideal Bose gas that typically results from the formation of a Bose-Einstein condensate (BEC) \citep{Reppy1984}. Since the Bose-Hubbard model describes an interacting Bose gas in a lattice that behaves as a superfluid below some critical temperature, one would expect superfluid \ce{^{4}He} in Vycor to conform to the Bose-Hubbard model behaviour. Indeed, this system is typically modelled using the XXZ model, which is a special case of the Bose-Hubbard model in the limit of large on-site interaction strength $U \gg 1$ \citep{van_otterlo_quantum_1995} including potentially interactions between photons on different sites. In this `hard-core boson' limit, no two bosons are allowed to occupy the same lattice site.

At large \ce{^{4}He} densities, a conventional phase transition between a superfluid phase and a Mott insulator phase is observed at finite temperature \citep{Fisher1989}. The critical temperature, $T_c$, at which this phase transition occurs decreases with the density $\rho$ of \ce{^{4}He}, reaching $T_c = 0$ at some positive density $\rho_c(T=0)$. At zero temperature, the system then undergoes a transition from a Mott-insulating state to a superfluid as the density $\rho$ crosses $\rho_c(T=0)$. In addition, the phase behaviour of \ce{^{4}He} adsorbed in Vycor exhibits an intermediate `Bose glass' phase, analogous to the intermediate phase of a type-II superconductor. Along with subsequent observations \citep{Weichman2008}, the quantum phase transition behaviour constitutes empirical evidence that the Bose-Hubbard model with density-dependent hopping and interaction parameters is a valid characterisation of the system within this parameter regime. As a result, this behaviour is structurally and formally similar to the zero-temperature superfluid-insulator phase transition of the \ce{^{87}Rb} atoms in the optical lattice.

\subsubsection{Triplons in Quantum Dimer Magnets}

Typical magnetic materials consist of an ordered arrangement of magnetic spin states. For `spin dimer compounds', pairs of spin states couple and, due to the crystalline structure of the material, interact only weakly with other coupled spin states. These weakly interacting `dimers', generate a paramagnetic `spin-liquid' ground state in the material comprised of local entangled spin singlet states, with an excitation gap to an excited triplet state. When a high strength magnetic field is applied to the material, Zeeman splitting of the triplet state closes the excitation gap, and the entangled spin singlets transition to the excited triplet state, and the material to a magnetically ordered state. 

The dimers in such systems behave as `bosonic quasiparticles' and, when excited by a magnetic field, are known as `triplons' \citep[1]{Nohadani2005}.  The phase transition from the paramagnetic phase to the ordered phase can be described as the formation of a BEC. In the appropriate parameter regime, the critical temperature of the transition vanishes, and so this phase transition is analogous to the zero-temperature transition from a Mott-insulating phase to a superfluid condensate. Moreover, the phase diagram of such quantum dimer compounds contains an intermediate partially polarised anti-ferromagnetic phase, making the phase behaviour of the material analogous to a type-II superconductor. This quantum phase transition behaviour has been verified experimentally \citep{Ruegg2003} and, moreover, is well modelled by a three-dimensional Heisenberg XY Hamiltonian, which can be derived from the Bose-Hubbard Hamiltonian in the hard-core boson limit $U \gg 1$ with no long-range interaction. We thus have empirical evidence that the Bose-Hubbard model describes the quantum dimer system within this parameter regime.

\subsubsection{Cooper Pairs in Josephson Junction Arrays}
\label{subsubsec:jjarray}

A Josephson junction array (JJA) is a granular superconductor given by an array of superconducting islands weakly coupled by Josephson tunnel junctions. The superconducting behaviour of the system is determined by the interplay between the strength of the coupling energy between the islands and the strength of the electrostatic interaction energy of Cooper pair charges at each island. High coupling energy between the islands leads towards high superconducting coherence. High interaction energy of Cooper pairs, controlled by the island capacitance, leads towards charge localisation on each island, and suppression of superconducting coherence \citep[569]{Bruder2005}. The behaviour of Josephson tunnelling and the interaction of Cooper pair charges is described by the quantum phase model Hamiltonian $H_{\text{QPM}}$ which is formally equivalent to the Bose-Hubbard Hamiltonian in the regime of large local particle number $\langle n_i \rangle \equiv \langle b_i^\dagger b_i \rangle \gg 1 $.

At high coupling energy, there is a critical temperature below which the array system is in a globally coherent superconducting state---the Cooper pairs `condense' into the same ground state. In the regime where the electrostatic interaction energy at each island is comparable to the coupling energy between adjacent islands, lowering the temperature of the array increases the resistance between islands, and the array undergoes a transition to an insulator phase. This phase transition is experimentally well explored \citep{vanderZant1996}, to the extent that the Bose-Hubbard model is taken as a valid characterisation of the behaviour of the JJA in a suitable parameter regime, with the Cooper pairs behaving as the bosons. As such, this phase transition is analogous to a zero-temperature superfluid--insulator phase transition in the optical lattice system.

\subsection{Summary and Prospectus}
\label{sec:summary analogue bh systems}

Each of the four systems discussed (i) is well described by the Bose-Hubbard model within an appropriate parameter regime, (ii) undergoes an analogue zero-temperature phase transition from an insulating phase to a superfluid phase, (iii) has a characteristic property that can be used to control the zero-temperature quantum phase transition, and yet (iv) has a distinctly different material constitution. The key details are summarised in \cref{tab:analogues}.  

\begin{table}
    \centering
    \begin{tabular}{llll}
    \toprule
     \it System & \it Boson &\it  Phase transition control & \it BH Parameters \\
    \midrule
    \midrule
    Cold atoms & \ce{^{87}Rb} atom & Magnetic field/laser properties \\
    \midrule
    \ce{^{4}He} adsorbed in Vycor & \ce{^{4}He} atom & \ce{^{4}He} density &  $U \gg 1, U_{i,j} \neq 0 $ \\
     Quantum dimer magnet & Dimer triplon & Magnetic field &  $U \gg 1$\\
     Josephson junction array & Cooper pair & Josephson energy and capacitance & $\langle n_i \rangle \gg 1 $\\
    \bottomrule
    \end{tabular}
    \vspace{1mm}
    \caption{Comparison of analogue Bose-Hubbard (BH) systems. The cold-atom system tuned to a certain parameter regime can serve as the source system to study various analogue target systems. Here, $U$ denotes the on-site interaction, $U_{i,j}$ the interaction strength between distinct sites $i$ and $j$, and $\langle n_i \rangle $ the expected value of the local particle numbers at site $i$.}
    \label{tab:analogues}
\end{table}

Ultimately, the purpose of exploring such systems is to learn more about type-II superconductors, with a view towards developing a better understanding of how they work and how we might be able to build, for instance, high-temperature superconductors. Some of the most promising naturally occurring candidates for such high-temperature superconductivity are so-called cuprate superconductors (materials characterised by alternating layers of copper oxides). Not only do these superconductors exhibit typical quantum phase transitions \citep{Zhou2022}, they are in fact best known for their remarkable magnetic behaviour, including the trapping of magnetic vortices in response to an external magnetic field. When these vortices are small enough (on the order of nanometers), as is the case in cuprate superconductors, the vortices exhibit ostensibly quantum behaviour \citep{Huebener2019}. However, these vortex states are difficult to observe directly, let alone probe experimentally \citep{Berthod2017}. Recent experiments employing the cold-atom source system above suggest that these vortex states can potentially be probed via analogue quantum simulation \citep{atala_observation_2014}---whether these experiments actually do probe such states is the subject of this work.

The cold-atom system can thus act as a versatile simulator of various Bose-Hubbard systems, since it is highly tunable and more effectively probed than the target systems. However, the model is only a good approximation of target system behaviour within some prescribed limit---that is, when the target systems are well described by, say, the quantum phase model, the XXZ model, or the XY model, which all reduce to the Bose-Hubbard model in a certain limit. Moreover, the cold-atom system exhibits behaviour that, given the right inferential structure, could enable the investigation of phenomena we think typical of type-II superconductors: quantum phase transitions between an insulator phase and superfluid phase, an intermediate phase between insulator and superfluid phases, and the quantum behaviour of magnetic vortex states generated by an external magnetic field.

In what follows we explore the nature of the inferential structure that would lead experimenters to have confidence that probing ultracold atoms in an optical lattice can tell them something about naturally occurring superconductors. For the three target systems above, we have experimental evidence for the phase transition in each system, which supports confidence that they are described by the Bose-Hubbard model in some limit. However, how might we gain confidence that we are successfully probing target-system behaviour when that behaviour falls outside the limits defined by the relevant model relations, such as quantum vortex states in cuprate superconductors? Answering such questions in general and specific circumstances will be the major occupation of the reminder of this paper.  

\section{Uniformity Principles in Analogue Quantum Simulation}
\label{sec:uniformity}

\subsection{Tokens, Types, and External Validation} 

An \emph{internally valid} experiment is one in which we genuinely learn about the source system we are manipulating. We will assume that all the experiments we consider here are internally validated by standard experimental means. To ensure that the outcomes of an experiment on a particular physical system are relevant to other physical systems with the same properties, we need to \emph{externally validate} the experiment. Typically, conventional experiments are performed with systems in mind that have the same, or a similar, material constitution. Such systems are believed to behave similarly when probed in the same circumstances. External validation then amounts to ensuring that the specific lab system has the same material properties as the target systems. More abstractly speaking, in an experiment a specific \emph{token} physical system is probed in order to learn about an entire \emph{type} of systems. The inference from the token to the type is based on a \emph{uniformity principle}, which asserts that all systems of the same material constitution behave in the same way when probed in the same circumstances. 

In analogue experiments, by contrast, scientists aim for a system of one type to stand in for a system of another type, the latter of which importantly has a distinct material constitution. 
In our case study, for instance, we have the source system consisting of cold atoms and the target systems consisting of a JJA or liquid Helium-4. It appears that \emph{by definition} we cannot make use of an intra-type uniformity principle between such systems since they are materially distinct. 

In order to establish that a system of one type can stand in for a system of another type, we would need to perform experiments on both systems in the same setting and compare their outcomes.
This would establish uniformity between tokens of different types. 
However, the purpose of an analogue quantum simulation is typically to probe the target system in a regime that is experimentally \emph{inaccessible}. How can we provide a reliable means for justification of the relevant chain of inferences in such circumstances? 
One way to achieve this would be to establish a specific \emph{inter-type uniformity principle} between certain systems.  
But how could inter-type uniformity be justified and which systems would fall under it? For intra-type uniformity principles, the criterion is clear: it is underpinned by the material constitution of the systems. For inter-type uniformity principles (even assuming their existence) this is less clear: 
Are we considering a uniformity principle between two types? Should all tokens of the type, in all parameter regimes, be captured by the uniformity principle? 

\subsection{Material Types, Empirical Types, and Universality Types}
\label{subsec:empvmat}

The natural implications of this discussion is that, in the context of analogue quantum simulation, we require uniformity principles that cut across the boundaries of different types. 
The corresponding notion of `type' is characterised by the material constitution of the systems. Let us therefore define the notion of a \emph{Material Type} as follows. 
\begin{description}
    \item[Material Type] Two token systems are of the same \emph{material type} if they share the same material composition, as determined by the properties and spatial arrangement of the constituent particles, atoms or molecules, at the relevant physical scale.  
\end{description}
This is a simple and intuitive notion of type in that it fleshes out the core conceptual idea of material sameness in a straightforward manner. Moreover, this idea of material type provides a simple and intuitive characterisation of the kind of uniformity principle that one might na\"ively take to underlie source--target inferences in a conventional experiment.  

To be applicable to real scientific examples there are of course a number of aspects of the idea of `material sameness' that need to be further specified---most obviously the \emph{scale} at which the material composition is required to be the same. To take a famous example, two samples of carbon atoms may be of the same material type at the atomic scale but of very different material types at the level of bonded allotropes. The project of characterising material types in a systematic and reliable manner will then crucially depend upon finding an appropriate scale of structure, be this atomic, molecular, mesoscopic, or even macroscopic. Furthermore, there are good reasons to think that material similarity alone cannot be sufficient to power the types of inferences made in experimental science. Consider the example of impurities. Clearly, whether such impurities in a source system are significant enough to render an inference between source and target systems unreliable depends on the form of inference and the sensitivity of the experimental protocol. It might be perfectly valid to treat two systems as of the same material type in the context of one experimental inference despite a high level of impurities in the target, say, but entirely invalid to treat the same two systems as of the same material type in the context of another experimental inference. 

The highly contextual nature of intra-type reasoning in experimental science might thus prompt us to reconsider the focus on material constitution as the basis for distinguishing types in an experimental context. At the very least, there is a strong motivation to move beyond a simplistic picture of experimental inference based upon source--target material similarity alone.\footnote{There are similarities here to the accounts of \cite{bursten:2018,roush:2018,norton:2021}.}

Our focus in what follows is on the specific structure of scientific inference in the context of analogue quantum simulation. We do not take ourselves to be articulating a view on experimental science in general. However, a possible first step towards such a view, motivated by the problems with the notion of material type, would be as follows. Plausibly, what matters in the context of an experimental inference is that the source and target physical systems should behave similarly in similar situations. Let us then define a notion of \emph{Empirical Type}. 
\begin{description}
    \item[Empirical Type] Two token systems are of the same \emph{empirical type}, in a specified parameter regime and with respect to a set of experimental prescriptions, if equivalent implementations of the  prescriptions in the parameter regime result in similar measurement outcomes.
\end{description}
We can thus understand the intra-type uniformity principles applied in conventional experimentation to be built around the assumption that tokens of a material type are also of the same empirical type. Such reasoning assumes that all tokens of the same material type can be described by a single theoretical model, which could then be validated by performing an experiment on a token system and applying the intra-material-type uniformity principle. To justify the use of such a principle, the experiment needs to be externally validated, which requires that the concrete token system we are probing is in fact representative of the type we want to make an inference about. For a material type, this amounts to establishing similarity in material constitution of the system. A similarity in \emph{nomic behaviour} is then assumed to lead to a similarity in empirical behaviour.

Whether or not this analysis is adequate in the context of conventional experimental science, it is clearly problematic in the context of analogue quantum simulation. In this context, scientists clearly are not aiming to justify an inference between source and target systems that are two tokens of the same material type, and are thus not looking to establish similarities in material constitution in order to establish nomic and empirical uniformity.

This is clear from the analogue systems we outlined in \S\ref{sec:case study}. Any putative inference from source to target in such simulations considers one physical material, say, \ce{^{87}Rb}, as a surrogate for another physical material, say, bound electron-phonon pairs in a superconductor or entangled spin states in a quantum dimer magnet. We might therefore seek to re-conceptualise the schema sketched above and consider an \emph{inter-material-type} uniformity principle that would underlie the reasoning at hand in place of the \emph{intra-material-type} uniformity principle. To make this explicit consider the idea of a \emph{Universality Type}:

\begin{description}
    \item[Universality Type] Two systems which are of different material types are of the same \emph{universality type} if, in a specified parameter regime and with respect to a set of experimental prescriptions, the behaviour displayed by the systems upon equivalent implementation of the prescriptions is appropriately similar and independent of differences in their material composition. 
\end{description}

A universality type is a particular kind of empirical type that additionally provides us with a potential route to external validation: an analogue quantum simulation might be validated on the basis of universality arguments showing the \emph{independence} of the measurement outcomes on material constitution between source and target. Such an argument would show that the source and target systems belong to the same universality type, and thus, other things being equal, would be of the same empirical type on that basis. In essence, inferential work previously performed by assumptions with regard to material types and laws of nature is now done by uniformity within the universality type. In each case the key step is to establish target and source as members of the same empirical type, but in the two cases this is achieved using a very different chain of reasoning.\footnote{This is a point of controversy in the literature: see \citep{Dardashti:2015,Dardashti:2019,Thebault:2016,EvansThebault2020} for the case in favour, and \citep{crowther:2019,Field:2022} for more sceptical commentary.} 

This suggests the question: can reasoning based on similarity as to empirical type be justified without appeal to material constitution \emph{or} universality arguments? In other words, can we justify uniformity principles between empirical types directly? 

\subsection{Empirical Quantum Types}

In this section we will consider a physical uniformity principle that cuts across material types based on independently established empirical evidence. This uniformity principle gains its inferential power by leveraging the validity of quantum theory in a well-characterised regime. 
The predictions of quantum theory have been confirmed to extremely high precision and at scales ranging from the size of the constituents of atoms to mechanical oscillators. In short, we are well justified to hold high confidence in the validity of quantum theory at the relevant scales, and knowledge of its applicability. 

As we will argue, this confidence in quantum theory can be used to justify a `quantum' notion of empirical type parallel to the idea of a universality type, the notion of an \emph{Empirical Quantum Type}. 
\begin{description}
    \item[Empirical Quantum Type] Two token systems are of the same \emph{empirical quantum type}, in a given parameter regime and with respect to given experimental prescriptions, if the same quantum mechanical model can be deployed in that parameter regime to provide an empirically reliable description of the systems for the experimental prescriptions.
\end{description}
Importantly, this definition allows for the possibility that the two systems at hand may be of different material types since, as explicitly illustrated by our case study, there clearly are cases in which the same quantum model can be employed to provide an empirically reliable description of systems with very different material constitution in the appropriate parameter regime. 

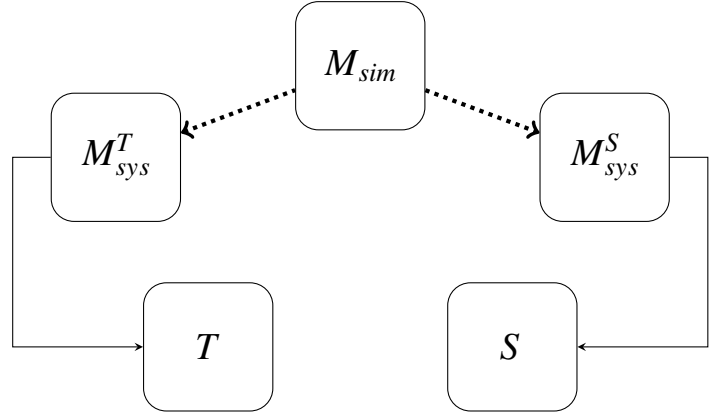
\begin{figure}[t]
  \begin{center}
    \large
    \begin{tikzpicture}[block/.style={draw=black,fill=white,rectangle,rounded corners=3mm,minimum width={17mm},minimum height=17mm}]
      \node [block] (S) {$M_{sim}$};
      \node [block] (B) [below left=-5mm and 15mm of S] {$M^{T}_{sys}$};
      \node [block] (E) [below right=-5mm and 15mm of S] {$M^{S}_{sys}$};
      \node [block] (F) [below left =8mm and -5mm of E] {$S$};
      \node [block,below right =8mm and -5mm of B] (A) {$T$};
      \draw [-stealth] (B.west) -- +(-0.5,0) |- (A.west);
      \draw [-stealth] (E.east) -- +(0.5,0) |- (F.east);
      \draw [ultra thick,black,dotted,<-] (B) to (S);
      \draw [ultra thick,black,dotted,->] (S) to (E);
    \end{tikzpicture}
  \end{center}   
  \caption{The inferential structure of analogue quantum simulation. The system models, $M^{T}_{sys}$ and $M^{S}_{sys}$, which we take to represent (solid arrows) the target $T$ and source $S$ systems, respectively, are de-idealisations in some controlled parameter regime (dotted arrows) of the simulation model $M_{sim}$ that defines the quantum empirical type.}

  \label{fig:aqs}
\end{figure}

More specifically, in the context of analogue quantum simulation, the relevant empirical quantum type is defined by an idealised \emph{simulation model} $M_{sim}$. For the simulation systems we outlined in \S\ref{sec:case study}, the simulation model is the Bose-Hubbard Hamiltonian $H_{\text{BH}}$. Given that both the target and the source physical systems are approximately described by the simulation model in a certain parameter regime, their empirical properties in this parameter regime will be approximately the same. Our most accurate description of the source and target systems will be specific \emph{system models}, $M_{sys}^S$ and $M_{sys}^T$, that include all known interactions and noise sources. Those are related to the simulation model by de-idealisation, see \cref{fig:aqs}. We can think of all tokens of an empirical quantum type that share the same material constitution, and therefore the same system model, as a material sub-type of the empirical type.

This way of thinking about an empirical quantum type in the context of analogue quantum simulation also provides a clear recipe for how to define what we called `equivalent experimental prescriptions' in \S\ref{subsec:empvmat}. 
We can specify an experimental prescription in terms of the idealised simulation model (say, $H_{\text{BH}}$) that jointly and approximately describes all tokens of the empirical quantum type (say, the cold-atom optical lattice and the JJA, each constrained to the appropriate parameter regime). 
In other words, as long as there is a well-defined way in which an experimental prescription can be specified and translated into equivalent prescriptions for systems of different material constitution, this prescription can figure in the definition of an empirical type. 
This prescription will often not be clear-cut, and will incorporate our understanding of the formal model, a qualitative understanding of the physical principles underlying the behaviour of the source and target systems, an understanding of how the applicability of these principles generates limitations on the parameter regime in which this behaviour arises, and an understanding of any contingencies of the specific experimental apparatus employed in the simulation. 
For tokens of different material sub-type, we can then exploit this understanding of our simulation systems to simultaneously de-idealise the experimental prescriptions in accordance with the de-idealisation to the respective system model, giving rise to equivalent (within some operational bound) experimental prescriptions.

\section{External Validation of Analogue Quantum Experiments}
\subsection{General Inferential Structure}
\label{subsec:inference}

Let us assume that we want to perform an experiment on a source quantum system $S$ in order to make an inference about another target quantum system $T$ of a distinct material constitution. Let us assume that the experiment on system $S$ is internally valid and thus that we have established that $S$ is accurately described by an idealised quantum simulation model $M_{sim}$ in the parameter regime~$P$. External validity in the context of an analogue quantum simulation experiment is equivalent to $S$ and $T$ being of the same empirical quantum type in \emph{the entire parameter regime relevant to the analogue quantum experiment}.

We can \emph{inductively} argue towards external validity as follows. Assume:
\begin{itemize}
    \item[(Q)] System $T$ is accurately described within the framework of quantum theory in a certain parameter regime $P$. 
    \item[(H)] System $T$ is accurately described by the idealised quantum simulation model $M_{sim}$ for some values of its parameters $P_0 \in P$.  
    \item[(R)] We have theoretical reasons to believe that $M_{sim}$ accurately represents $T$ in the parameter regime~$P$.
\end{itemize}
We can then \emph{inductively infer} that:  
\begin{itemize}
    \item[(C)]  System $T$ is accurately described by $M_{sim}$ in the entire parameter regime~$P$. 
\end{itemize}
Since $S$ and $T$ are taken to be of the same empirical quantum type in the entire parameter regime relevant to the analogue quantum simulation, the experiment is then externally valid.

Condition (Q) is supported by our confidence in the empirical reliability of quantum theory as a whole within a given parameter regime. Contemporary physics provides us with a wealth of evidence regarding the systems and regimes in which quantum behaviour will be found. This evidence is wide and varied, including experimental evidence from more than a century of manipulating a broad range of quantum systems, and theoretical evidence from powerful frameworks, such as effective field theory, that provide us with considerable confidence that we understand the relevant scales at which quantum theory can be applied.

Condition (H) is established by conducting a conventional experiment on the target system or a token of the same material sub-type (in the conventional sense). While the target system $T$ may be inaccessible in some parameter regime of interest, it is typically accessible in some other regime that can be experimentally probed. Moreover, given that we are in the realm of applicability of quantum theory, in this regime we also want to be able to compare the predictions of quantum theory with experimental outcomes, so it is advantageous to perform an experiment in the computationally tractable regime.

Despite the fact that the target system $T$ may be inaccessible in some parameter regime of interest, our expectation that the simulation model could well apply in this regime is captured by condition (R). This expectation is underpinned by our confidence that quantum theory is the right modelling framework for the relevant scale and empirical context, and promotes the belief that the system will exhibit the relevant model behaviour in the broader regime.

We can formulate a specific claim based on this argument pattern as follows:\footnote{We can express Claim 1 in Bayesian terms as $P[C| Q + H + R] - P[C]>0$ where C, Q, H, R are the values of propositional variables corresponding to the truth/falsity of the relevant claims, $P[A|B]$ is the conditional probability of $A$ given $B$, and we have assumed non-trivial prior probabilities.}
\begin{itemize}
\item[] \textbf{Claim 1}: Assumptions (Q), (H), and (R) jointly provide inductive support for the conclusion (C) such that learning the conjunction $\text{(Q)} \wedge \text{(H)} \wedge \text{(R)}$ gives defeasible justification for raising ones degree of belief in (C).
\end{itemize}
The reasoning behind Claim 1 is a form of `bootstrapping' argument which allows us to extend the parameter range in which we can have confidence that $S$ and $T$ are the same empirical type. At its core, the bootstrapping argument towards external validity works by leveraging a small piece of empirical knowledge of the target system in a narrow regime as captured by condition (H) in order to generalise the applicability of the model $M_{sim}$ to a broad parameter regime based on the quantum uniformity principle (Q). The condition (Q) is a key epistemic tool for the external validation of an analogue simulation, because it buttresses the inferential connection between the quantum behaviours displayed by $S$ and $T$.  Given (Q), it is sufficient to validate the simulation model for \emph{specific} parameters and inductively extend the applicability of the model beyond those parameters to the broader regime $P$, so long as we have theoretical reasons to believe that the simulation model is still applicable in the broader regime (R). Condition (R) is necessary here since the behaviour of $T$ in the parameter regime $P$ is empirically inaccessible, and so such \emph{theoretical} reasons are often the only evidence we have to the behaviour of $T$ in $P$. 

Our appraisal of the inferential situation has two further significant consequences with respect to the \emph{degree} of inferential support that the package $\text{(Q)} \wedge \text{(H)} \wedge \text{(R)}$ gives in comparison to alternative reasoning patterns which rely only on a subset of the inductive premises. We can set out these implications as follows:\footnote{In Bayesian terms Claim 2 is equivalent to $P[C| Q + H + R] \gg P[C| H + R]$ and Claim 3 is equivalent to $P[C| Q + H + R] \gg P[C| Q + R]$ where the $\gg$ sign should be read qualitatively as indicating that the inductive support is nontrivially larger, but not necessarily many orders of magnitude larger.  We make no claims regarding the scale of these values of inductive support, and it is perfectly plausible that they be low relative to the inductive support that may accrue in a conventional pattern of experimental inference. Plausibly, however, we take our arguments to imply that one may understand $P[C| Q + H + R] - P[C]$ to be nontrivial even if one expects trivial inductive support in cases where the target system is entirely inaccessible, and thus that $P[C| Q + R] - P[C] \approx 0$, c.f. \cite{Dardashti:2019,Field:2022}.} 
\begin{itemize}
\item[] \textbf{Claim 2}: The \emph{degree} of inductive support for the conclusion (C) provided by assumptions (Q), (H), and (R) is non-trivially greater than that provided by (H) and (R) alone.
\end{itemize}
\begin{itemize}
\item[] \textbf{Claim 3}: The \emph{degree} of inductive support for the conclusion (C) provided by assumptions (Q), (H), and (R) is non-trivially greater than that provided by (Q) and (R) alone.
\end{itemize}
To see that this is the case, consider the inferential weakness of reasoning based upon the relevant subsets of premises. 

With respect to Claim 2, consider a situation in which we assume (H) and (R) but not (Q). In such circumstances we have \emph{experimental} evidence that $T$ is accurately described by the idealised quantum simulation model $M_{sim}$ in a specific parameter regime and we have \emph{theoretical} reasons to believe that $M_{sim}$ accurately represents $T$ in the wider parameter regime. However, without (Q) we have no inferential link between the behaviour of $T$ and the behaviour of $S$, and therefore no link to the (by assumption internally valid) experiment which probes $S$ in the salient regime. Condition (Q) captures the well-justified assumption that the modelling framework of quantum theory, as provided by the core apparatus of Hilbert space representation together with some minimal interpretation given by the Born rule, does not break down (and is almost certainly applicable) at the relevant scales at which, for instance, quantum vortex states arise in type II superconductors. Put simply, based on our general empirical knowledge about the applicability of quantum mechanics, it is a very reasonable assumption that the elementary objects at play in the source and target systems of an analogue quantum simulation are quantum objects. Without (Q), on their own (H) and (R) provide a comparatively weak inductive base for the conclusion (C) precisely because the relevant bootstrapping argument can get no traction. This supports Claim 2.

With respect to Claim 3, consider a situation in which we assume (Q) and (R) but not (H). In such circumstances there \emph{is} an inferential connection between the behaviour of $S$ and $T$. However, since our assumption no longer contains any \emph{experimentally derived} inductive evidence regarding the system $T$, the strength of inference we can make is greatly diminished. In particular, we are open to the possibility that beyond the features encoded in our broad quantum uniformity hypothesis, our basic theory of the target system may be completely wrong. Without (H) we have no \emph{specific empirical evidence} that guides the selection of the most adequate model within the modelling framework of quantum theory to describe the target. All we have to constrain our reasoning with respect to the target system is theory, and although (Q) gives us a principle to connect $S$ and $T$, it does not alone licence strong reasoning with regard to the detailed physics underlying $T$. Hence Claim 3 is supported.\footnote{We note that Claim 3 is very much in line with the analysis of  \citet{Field:2022} of the case of analogue simulations in which the target system is inaccessible and the relevant inferential link is built in terms of universality arguments, c.f. \citep{crowther:2019}.}

Let us now consider the specific implementation of this novel yet robust pattern of inference in the context of our case study.

\subsection{External Validation of Bose-Hubbard Analogue Simulations}
\label{subsec:bhvalidation}

Our framework for understanding the external validity of analogue quantum experiments can be applied to the context of the Bose-Hubbard analogue simulations outlined in \S\ref{sec:case study}. The optical lattice system is our source system $S_{CA}$, and the superfluid helium, quantum dimer magnet, and JJA are our target systems $T_{\text{He}}$, $T_{\text{DM}}$, $T_{\text{JJA}}$, respectively. We take each of these systems to be described by the same idealised quantum simulation model $M_{sim}$---the Bose-Hubbard model $H_{\text{BH}}$---in the right parameter regime $P_{0}$ (such as for $U \gg 1$ or $\langle n_i \rangle \gg 1$). And we take there to be high confidence that the relevant probing experiments on the optical lattice system are internally valid.

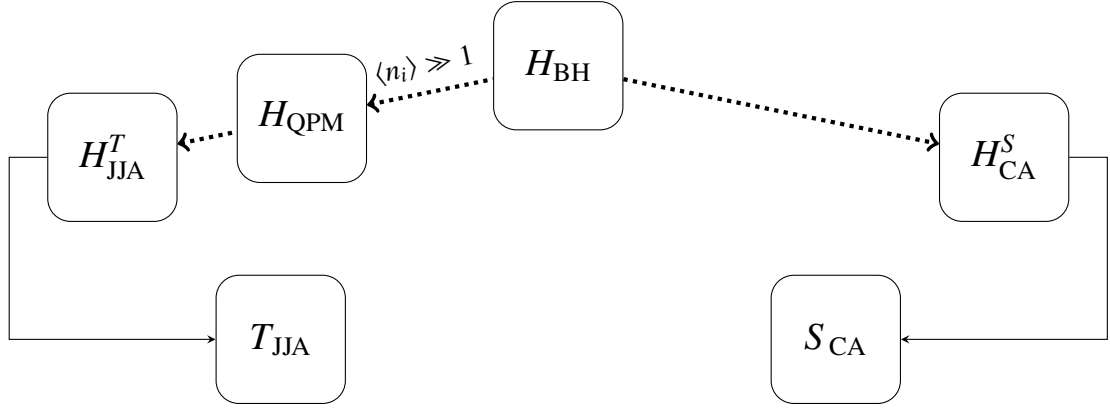
\begin{figure}[t]
  \begin{center}
    \large
      \begin{tikzpicture}[block/.style={draw=black,fill=white,rectangle,rounded corners=3mm,minimum width={17mm},minimum height=17mm}]
      \node [block] (S) {$H_{\text{BH}}$};
      \node [block] (B) [below left=-5mm and 4.15cm of S] {$H_{\text{JJA}}^{T}$};
      \node [block] at ($(S)!.575!(B)$) (QPM) {$H_{\text{QPM}}$};
      \node [block] (E) [below right=-5mm and 4.15cm of S] {$H_{\text{CA}}^{S}$};
      \node [block] (F) [below left =7mm and 5mm of E] {$S_{\text{CA}}$};
      \node [block,below right =7mm and 5mm of B] (A) {$T_{\text{JJA}}$};
      \draw [-stealth] (B.west) -- +(-0.5,0) |- (A.west);
      \draw [-stealth] (E.east) -- +(0.5,0) |- (F.east);
      \draw [ultra thick,black,dotted,<-] (B) to (QPM);
      \draw [ultra thick,black,dotted,->] (S) to (E);
      \draw [ultra thick,black,dotted,<-] (QPM) to  node [sloped, above, midway](M)  {\small $ \langle n_i \rangle \gg 1$} (S);
      
    \end{tikzpicture}
  \end{center}   
  \caption{The inferential structure of the analogue quantum simulation of a JJA system, $T_{\text{JJA}}$, by a cold-atom optical lattice system, $S_{\text{CA}}$. As discussed in \S\ref{subsubsec:jjarray}, the JJA system is described by the quantum phase model Hamiltonian $H_{\text{QPM}}$ which can be reduced to the Bose-Hubbard model in the limit $\langle n_i \rangle \gg 1$.}
  \label{fig:jjcoldatom}
\end{figure}

To see how this schema works in practice, let us consider an example. Although the ultimate goal of such quantum simulations is to learn about, say, the nature of natural type-II superconductors, it will be instructive to consider as an example the JJA target system, $T_{\text{JJA}}$. The inferential structure of this simulation is depicted in \cref{fig:jjcoldatom}. According to our framework, whether the optical lattice analogue quantum simulation counts as externally valid turns on whether $T_{\text{JJA}}$ each of (Q), (H), and (R) are satisfied.

Beginning with condition (H), we considered above in \S\ref{subsubsec:jjarray} the manner in which the superconductor--insulator transition is established by way of conventional experimentation on the JJA. It is illustrative, however, for understanding the inferential role played by the narrow parameter regime $P_{0}$ to provide some more detail. Suppose we are interested in determining the critical value of the ratio $x_{cr}$ of the Josephson coupling energy to the capacitance at which we think a superconductor--insulator phase transition occurs in a JJA, and so support the claim that the Bose-Hubbard model is the appropriate model for the system. Analytic and quantitative investigations using the Bose-Hubbard model provide a zero-temperature hypothesis for where this value might lie. Observations of the array show that, for each trialled value of $x_{cr}$, there is a characteristic response in the resistance across the array as a function of lowering the temperature. Graphical inspection of these characteristic responses enables the determination of the boundary between superconducting and insulating behaviour, and so also a determination of the value of $x_{cr}$. Comparison of this value with analytic derivations lends support to the proposal of a quantum phase transition in the system \citep{Fazio2001}. Without this direct empirical evidence of the applicability of $H_{\text{BH}}$ to $T_{\text{JJA}}$ in the narrow parameter regime, we would be incapable of assuming (H), and so be in a relatively impoverished situation with regards to inductively supporting (C), as per Claim 3.

However, a number of idealisations are required to enable the phase transition to emerge, and not only to ensure that the system is well characterised by the Bose-Hubbard model. In practice, the dynamical behaviour of the array can be influenced by random offset charges at each island, which introduce an intrinsic degree of disorder to the array, especially at the phase transition boundary, dissipation due to coupling to the environment, which can dampen coherence effects across the array, and the creation of quasiparticles, which have been unexpectedly detected at milliKelvin temperatures, which exacerbate dissipation effects \citep{vanderZant1996}. As such, quantifying by way of direct conventional experimentation the nature of the superconductor--insulator phase transition outside of the parameter regime where these effects are negligible ($P_{0}$) is very difficult and, at certain fine-grains, essentially impossible. But we might still have an expectation that the system can be characterised by $H_{\text{BH}}$ outside of this constrained parameter regime $P_{0}$. In particular, we might expect that JJAs, as granular superconductors, will admit magnetic vortex states that display quantum dynamics.

This expectation is captured by condition (R). Upon establishing that an analogue of a zero-temperature quantum phase transition is occurring in the JJA system characterised by $H_{\text{BH}}$, our knowledge of general Bose-Hubbard systems then implies that such behaviour will be exhibited in a broader regime, one which is inaccessible to probing by conventional experiment due to the complexity or intractability of the system in that regime.

Condition (Q) is established independently of the analogue simulation, and is the key to the external validity of analogue \emph{quantum} experiments. There is a multitude of independent lines of evidence that superconductors, and so JJAs, are well described by quantum theory, and perhaps even likely in a parameter regime much broader than $P$, but certainly within the parameter regime set by the limits of what can be probed by the cold-atom optical lattice source system. The interplay between theory and experiment that has allowed us to be relatively confident that Cooper pairs, and their behaviour as bosons in superconductor--insulator phase transitions, can be described in the modelling framework of quantum theory reduces the inferential burden on external validation in analogue quantum simulations. Without this independent evidence of the quantum behaviour of superconductors, we could not assume (Q), and so would again be in a relatively impoverished situation with regards to inductively supporting (C), as per Claim 2.

This example demonstrates that the inferential structure of analogue quantum simulation relies on a kind of consilience between (Q), (H), and (R): (Q) sets the general empirical foundation on which we can use (R) to obtain a specific theoretical basis to support conclusions regarding the detailed physics of $T$, while (H) provides a more narrow experimental basis to support claims regarding the dynamical behaviour of $T$. Although there may be phenomena in $T$ that we are unable to probe or manipulate experimentally---such as quantum vortex states---we take superconductors to be well-understood systems within a prespecified range of scales established by a lengthy tradition of interplay between superconductor theory and experiment. Moreover, there is a sense in which the appropriate parameter regime in system $T$ is being set by our knowledge and experience probing relevant phenomena in the source system $S$. We observe some phenomenon in $S$, like a quantised magnetic vortex state, only under a particular set of conditions, and we expect there to be an analogue set of conditions in $T$. This expectation is underpinned by the consilience between (Q), (H), and (R).

The combination of (Q) + (H) + (R) then allows us to argue inductively that the target system $T_{\text{JJA}}$ is described by $H_{\text{BH}}$ in a parameter regime $P$ that is broader than the regime in which we have direct conventional empirical evidence ($P_{0}$). More specifically, it is the combination of these three conditions that provides the relevant inductive base for inferences about properties of inaccessible concrete phenomena in JJAs, such as quantum vortex states, based on the observation of such phenomena in the cold-atom system. Because we can validate the applicability of Bose-Hubbard dynamics in the target systems in some tractable regime, we are justified in making inferences about the Bose-Hubbard behaviour of those systems in intractable regimes based on the behaviour observed in the analogue simulation experiments. This inference underpins the claims typical of analogue quantum simulations that probing the accessible behaviour of the source system can be taken to probe the inaccessible behaviour of the target system.

Such arguments in effect justify simultaneous de-idealisation of a single abstract quantum model to source and target system within a designated range of applicability, and in so doing provide justification for reasoning based upon regularity within empirical types. As we noted above, the de-idealisation of the simulation model to the respective system models, which in practice establishes the operational prescriptions that underpin regularity across empirical types, is not particularly clear cut. In short, it is our practical understanding of the physical systems, the nature of our models, the physical principles that underlie those models, how these constrain the parameter regimes within which they are applicable, and the contingencies of our empirical access to the respective systems in the laboratory that all play a key role in de-idealisation. We intend much of our discussion above of such practical understandings and contingencies to provide a guide to the architecture of this process.

There are some caveats here, of course. It is important to note that there are limitations on the applicability of (Q)---that is, there are limitations on the regime in which the JJA will be accurately described by quantum theory. At a certain level of coarse-grained abstraction, the JJA will behave classically. We do not expect the analogue quantum simulation to provide evidence for behaviour in this extended parameter regime. But at the appropriate fine-grained description---at which one can generate confidence in the quantumness condition (Q)---we can then infer the relevant Bose-Hubbard model to be a suitable description.

We thus reach the remarkable conclusion that although the JJA and the cold-atom optical lattice are instances of wholly different material constitutions, we expect them to obey structurally similar phase space and critical point dynamics on account of the strength of the analogue simulation: an inter-type uniformity principle becomes an intra-empirical-quantum-type uniformity principle, which then underpins the external validity of the analogue experiments.

\section{Conclusion}
\label{sec:discussion}

This paper has provided the first philosophical investigation of the epistemology of the novel experimental scientific practice of analogue quantum simulation. 
This practice can be understood as `wavefunction engineering' since it relies on both systems exemplifying the same empirical quantum type: 
despite having different material constitution, the same quantum wavefunction can be used to accurately represent both systems in a relevant regime.
We have argued that, in such contexts, limited empirical access to both source and target system can be leveraged to external validation of analogue simulations by the independently and empirically established confidence in the validity of quantum theory in both systems. Crucial here is appeal to a quantum uniformity principle that can be understood as a meta-principle for modelling practice. As the practice of wavefunction engineering and analogue experimentation continues to thrive the form and strength of such patterns of inference will become increasingly relevant to scientific practice and thus, we trust, to the philosophy of the scientific method.

One might wonder where the bulk of the work is done in this argument: On the one hand there is the underlying, but broad, uniformity principle, and on the other hand there is the specific, but narrow, empirical evidence for the validity of the model due to direct observation. Specifically, one might ask whether the uniformity principle is adding a quantitative or a qualitative difference to the argument. After all, it is standard practice to confirm models by performing experiments in restricted parameter regimes. We argue that the difference is qualitative: we would not be able to conclude the broad validity of the simulation model in \emph{both} source and target system across the entire parameter regime of interest if we were not very confident in the validity of the modelling framework.

And indeed, there are prominent examples of analogue experimentation in which we do not have empirical evidence that the modelling framework is adequate for the target system. In particular, this is the case in the context of remote or entirely inaccessible phenomena such as analogue gravity \citep{Dardashti:2015}, wherein justificatory arguments are framed in terms of the universality of phenomena across different material types. It remains to be seen, however, how strongly the distinction between such cases and cases like those we have considered should be taken. On the one hand, as argued by \cite{winsberg:2010} in the context of experimentation and classical computer simulation, if we want to characterize the difference between two methods we should not focus on what objective relationship actually exists between the object of an investigation and its target. Rather, what distinguishes different methods is the character of the argument given for the legitimacy of the inference from object to target and the character of the background knowledge that grounds that argument. On such a view the distinction between wavefunction engineering and analogue experimentation based on universality would be a robust one as the type of argument to support the inference is distinct. However, on the other hand, at an ontological level, the distinction between \emph{intra}-empirical-type uniformity and inter-material-type uniformity is not grounded in a clean or straightforward distinction. 

Similarly, we can compare analogue simulation to both standard experimentation and simulation. Taking again Winsberg's view as the basis, the distinction between simulation and experimentation is grounded in what kind of evidence we refer to when justifying inferences. One could consider a speculative thesis, worthy of future consideration, along these lines as follows. First, one might think that arguments for the validity of a computer simulation are \emph{model-based}, whereas arguments for the validity of an experiment are \emph{nomology-based}. Then, second, given our empirical-quantum-type argument, analogue quantum simulation could be taken to be a practice that is genuinely intermediate between simulation and experimentation. Its justification is grounded simultaneously in both a model-based simulationist and nomology-based experimentalist reasoning. Third and finally, we would then have that a model-based epistemology and nomology-based ontology of simulation and experimentation cannot be separated. Rather, since it is the mode of de-idealisation that is different in the two cases, we should not be trying to differentiate between what there is and what we know. As the practice of wavefunction engineering and analogue experimentation continues to thrive, such issues will become of increasing importance, and thus warrant further investigation.

\section*{Acknowledgements}

We wish to thank Ivan Kassel, Gerard Milburn, and Andrew White for illuminating discussions on this topic. We also thank the anonymous referees for their helpful comments. We are also grateful to the audience of the Symposium `From Bosons to Markets to Black Holes: New Prospects for Analogical Reasoning' at the PSA 2021 meeting in Baltimore. P.W.E. acknowledges support from the University of Queensland and the Australian Government through the Australian Research Council (DE170100808) and from the Foundational Questions Institute and Fetzer Franklin Fund (FQXi-RFP-CPW-2019), a donor advised fund of Silicon Valley Community Foundation. D.H.\ acknowledges financial support from the U.S.\ Department of Defense through a QuICS Hartree fellowship.

\end{document}